\documentclass[11pt]{article}
\newcommand{\Comment}[1]{{}}
\usepackage[textwidth = 430 pt, textheight = 630 pt]{geometry}
\usepackage{amssymb,euscript,amsmath,amsfonts}

\usepackage{color}
\definecolor{MyDarkBlue}{rgb}{0.15,0.15,0.45}
\usepackage[linktocpage=true]{hyperref}
\hypersetup{
colorlinks=true,
citecolor=MyDarkBlue,
linkcolor=MyDarkBlue,
urlcolor=MyDarkBlue,
pdfauthor={Neil Lambert, Constantinos Papageorgakis and Maximillian Schmidt-Sommerfeld},
pdftitle={Instanton Operators in Five-Dimensional Gauge Theories},
pdfsubject={hep-th}
}

\usepackage[numbers,sort&compress]{natbib}
\usepackage{hypernat}

\newcommand\ignore[1]{}
\def\one{{\,\hbox{1\kern-.8mm l}}}

\def\Tr{{\rm Tr\, }}

\newcommand{\SO}{\mathrm{SO}} 
\newcommand{\SU}{\mathrm{SU}} \newcommand{\U}{\mathrm{U}}

\newcommand{\Cset}{{\,\,{{{^{_{\pmb{\mid}}}}\kern-.45em{\mathrm C}}}}}

\newcommand{\cI}{\mathcal I}

\newcommand{\cN}{\mathcal N}\newcommand{\cO}{\mathcal O}

\newcommand{\ie}{{\it i.e.\;}}
\newcommand{\eg}{{\it e.g.\;}}
\newcommand{\be}{\begin{equation}}
\newcommand{\ee}{\end{equation}}
\newcommand{\bea}{\begin{eqnarray}}
\newcommand{\eea}{\end{eqnarray}}

\begin{document}

\renewcommand{\thefootnote}{\fnsymbol{footnote}}

\rightline{MTH-KCL/2014-20}
 \rightline{QMUL-PH-14-25}

   \vspace{1.8truecm}

 \centerline{\LARGE \bf {\sc Instanton Operators in}}
 \vskip 12pt
 \centerline{\LARGE \bf{\sc 
 Five-Dimensional Gauge Theories}   }

\centerline{\LARGE \bf {\sc  }} \vspace{2truecm} \thispagestyle{empty} \centerline{
    {\large {\bf {\sc N.~Lambert${}^{\,a,}$}}}\footnote{E-mail address: \href{mailto:neil.lambert@kcl.ac.uk}{\tt neil.lambert@kcl.ac.uk}}{,}
    {\large {\bf{\sc C.~Papageorgakis${}^{\,b,}$}}}\footnote{E-mail address:
                                 \href{mailto:Costis Papageorgakis
                                   <c.papageorgakis@c.papageorgakis@qmul.ac.uk>}{\tt
                                   c.papageorgakis@qmul.ac.uk}} {and}
                               {\large {\bf{\sc M.~Schmidt-Sommerfeld${}^{\,c,d,}$}}}\footnote{E-mail address:
                                 \href{mailto:Maximilian Schmidt-Sommerfeld <maximilian.schmidt-sommerfeld@hotmail.com>}{\tt maximilian.schmidt-sommerfeld@hotmail.com} }
                                                           }

\vspace{1cm}
\centerline{${}^a${\it Department of Mathematics}}
\centerline{{\it King's College London, WC2R 2LS, UK}}
\vspace{.4cm}
\centerline{${}^b${\it CRST and School of Physics and Astronomy}}
\centerline{{\it Queen Mary University of London, E1 4NS, UK}}
\vspace{.4cm}
 \centerline{${}^c${\it Arnold-Sommerfeld-Center f\"ur Theoretische Physik, LMU M\"unchen}}
 \centerline{\it Theresienstra\ss e 37, 80333 M\"unchen, Germany}
 \vspace{.4cm}
 \centerline{${}^d${\it Excellence Cluster Universe, Technische Universit\"at M\"unchen}}
 \centerline{\it Boltzmannstra\ss e 2, 85748 Garching, Germany}
 
\vspace{.5truecm}

 
\thispagestyle{empty}

\centerline{\sc Abstract}
\vspace{0.4truecm}
\begin{center}
\begin{minipage}[c]{360pt}{
    \noindent We discuss instanton operators in five-dimensional
    Yang-Mills gauge theories. These are defined as disorder operators
    which create a non-vanishing second Chern class on a four-sphere
    surrounding their insertion point. As such they may be thought of
    as higher-dimensional analogues of three-dimensional monopole (or
    `t Hooft) operators. We argue that they play an important role in
    the enhancement of the Lorentz symmetry for maximally
    supersymmetric Yang-Mills to $\SO(1,5)$ at strong coupling.
}

\end{minipage}
\end{center}
 
\vspace{.4truecm}

\noindent

\vspace{.5cm}

\setcounter{page}{0}

\renewcommand{\thefootnote}{\arabic{footnote}}
\setcounter{footnote}{0}

\linespread{1.1}
\parskip 4pt

{}~
{}~

\makeatletter
\@addtoreset{equation}{section}
\makeatother
\renewcommand{\theequation}{\thesection.\arabic{equation}}

\newpage

\section{Introduction}

One of the more dramatic results to come out of the study of strongly
coupled string theory and M-theory was the realisation that there
exist UV-complete quantum super-conformal field theories (SCFTs) in
five and six dimensions \cite{Witten:1995zh, Seiberg:1996bd,
  Morrison:1996xf, Seiberg:1997ax, Intriligator:1997pq}. These
theories then provide UV completions to a variety of perturbatively
non-renormalisable five-dimensional (5D) Yang-Mills theories. In this
paper we will consider the notion of `instanton operators' (or Yang
operators) and explore their role in five-dimensional
Yang-Mills. These local operators are a natural higher-dimensional
analogue of monopole (or 't Hooft) operators in three dimensions
\cite{'tHooft:1977hy, Borokhov:2002ib, Borokhov:2002cg} which are \eg
important in including eleven-dimensional momentum transverse to
M2-branes in the ABJM model \cite{Aharony:2008ug}. Indeed, in a flux
background M2-branes expand into M5-branes and the magnetic flux of
the monopoles is mapped into instanton flux
\cite{Lambert:2011eg}. Therefore one can expect an operator similar to a monopole operator in 3D to
play as important a role for M5-branes, that is in the relationship
between the six-dimensional (2,0) SCFT and five-dimensional
Yang-Mills.

5D Yang-Mills has a conserved current
\begin{align}
  \label{eq:10}
 J = \frac{1}{8 \pi^2} \Tr \star (F \wedge F) 
\end{align}
and the instantons of the theory are BPS particles, also referred to
as instanton-solitons, which carry the associated charge of
$J$. Intuitively, instanton operators act as
`instanton-soliton-creating operators', which insert a topological
defect at a spacetime point. This imposes certain (singular) boundary
conditions for the behaviour of the gauge field at the insertion
point. The classical equations compatible with such a structure were
first solved by Yang in the 70's \cite{Yang:1977qv}, in the context of
generalising the Dirac monopole solution to a static, $\SO(5)$
symmetric particle in 6D $\SU(2)$ gauge theory. This solution, being
time-independent, is a solution of (Euclidean) 5D super-Yang-Mills and
can be extended to any gauge group. These turn  out to be
$\frac{1}{2}$-BPS only for $\rho=0$, where the $\rho$ denotes the
instanton size modulus.

More formally, instanton operators are defined in a manner familiar
from three dimensions: they modify the boundary conditions for the
gauge field in the path integral of 5D Yang-Mills
theories. An immediate objection may be that naively we cannot define
a theory by a non-renormalisable lagrangian and consequently we also
cannot define an operator by a path integral prescription based on
such a lagrangian. However, one may appeal to the conjecture of
\cite{Douglas:2010iu,Lambert:2010iw} that the maximally supersymmetric
5D Yang-Mills lagrangian is non-perturbatively UV
complete (\eg along the lines of \cite{Papageorgakis:2014dma}) and
does define a theory. In such a scenario, instanton operators may play
a crucial role in   the UV completion, resulting in a
self-consistent picture.\footnote{For other attempts to formulate the
  $(2,0)$ theory, which have an explicit dependence upon the extra
  dimension, see for example \cite{Samtleben:2011fj,
    Chu:2011fd,Bonetti:2012st, Saemann:2012uq, Ho:2014eoa,
    Buchbinder:2014sna}.}  In any case, this definition extends to any
five-dimensional theory in which there is a notion of a gauge field
strength and therefore we expect that instanton operators can be
extended to any formulation.

Along these lines, we will give evidence for how these local operators
can be used in the maximally supersymmetric case to insert discrete
units of six-dimensional momentum. As such they are crucial for the
Lorentz symmetry enhancement to $\SO(1,5)$ at strong coupling. Our
discussion also touches upon the interesting topic of compactifying
CFTs in the absence of a lagrangian
description, and what this implies for the correlation functions. In particular, we hope to shed some light on how a 6D CFT
with no free parameters or marginal operators can be related to an
interacting 5D Yang-Mills theory.

The rest of this note is organised as follows. In Section
\ref{instops} we define instanton operators and explore some of their
elementary properties. In Section \ref{applic} we then apply them to
the case of maximally supersymmetric 5D Yang-Mills and show how they
can lead to Lorentz symmetry enhancement. We briefly conclude in
Section~\ref{summary}.

\section{Instanton Operators}\label{instops}

We will define an instanton operator in analogy with monopole
operators in three dimensions. In particular, we consider the
Euclidean regime of the theory consisting of gauge fields $A$, scalars
$X$ and fermions $\psi$ where
\begin{align}
\langle{\cal O}_{01}(x_1)\ldots {\cal O}_{0k}(x_k)\rangle = \int [DXDAD\psi] \;{\cal O}_{01}(x_1)\ldots {\cal O}_{0k}(x_k) e^{-S}\;,
\end{align}
with the $\cO_{0i}(x_i)$ some local five-dimensional operators. We then
introduce a new local operator ${\cal I}_n(x)$, which modifies the
boundary conditions of the gauge field at infinity via the condition
\begin{align}
\langle{\cal I}_n(x){\cal O}_{01}(x_1)\ldots {\cal O}_{0k}(x_k)\rangle = \int_{
\frac{1}{8\pi^2}{\rm Tr}\oint_{S^4_x} {F\wedge F}=n} [DXDAD\psi]\; {\cal O}_{01}(x_1)\ldots {\cal O}_{0k}(x_k) e^{-S}\ ,
\end{align} 
where $S^4_x$ is an arbitrary four-sphere that surrounds the point
$x\in {\mathbb R}^5$.  In other words, inserting ${\cal I}_n(x)$ into
a correlator instructs us to integrate over field configurations that
carry non-vanishing instanton number on the four-sphere surrounding
the insertion point.

First, we need to check that the fields obey the classical equations
of motion near the insertion point. In fact, the classical solution
corresponding to a single $\SO(5)$-symmetric instanton operator for
$\SU(2)$ gauge group was considered long ago in \cite{Yang:1977qv} by
Yang, where it appears as a static soliton in six dimensions. A
stringy embedding of the $\SU(N)$ generalisation was given later by
Constable, Myers and Tafjord \cite{Constable:2001ag}, in the context
of D1$\perp$D5 intersections.

To construct such a solution one introduces spherical coordinates on Euclidean ${\mathbb R}^{5}$:
\begin{align}\label{coords}
  ds^2 &= \delta_{\mu\nu}dx^\mu dx^\nu\cr
   &= dr^2 + r^2\gamma_{ij}d\theta^i d\theta^j\;,
\end{align}
with $\gamma_{ij}$ the metric on the four-sphere. We wish to solve the
Yang-Mills equations
\begin{align}
 D^\mu F_{\mu\nu}
=0\;,\qquad D_{[\mu}F_{\nu\lambda]} =0\;, 
\end{align}
but with non-vanishing 
\begin{align}\label{instnumb}
I = \frac{1}{8\pi^2} {\rm Tr}\oint F\wedge F\;, 
\end{align}
over any given sphere of radius $r$ centred about the origin.

Note that since
\begin{align}
 d~{\rm Tr}(F\wedge  F ) =0\;,
\end{align}
we see that if $I$ is evaluated on two spheres of radii $r_1$ and
$r_2$ then 
\begin{align}
 {\rm Tr}\oint_{S_1}F\wedge F- {\rm
  Tr}\oint_{S_2}F\wedge F = \int_{B_{12}}  d ~{\rm  Tr}(F\wedge F )=0\ ,
\end{align}
 where $B_{12}$ is the `annulus' region whose boundary is the two
spheres of radii $r_1$ and $r_2$. Thus the topological charge $I$ is
constant along the radial direction. Since the size of the sphere grows as
$r^4$ we have that 
\begin{align} 
F_{\mu\nu}\sim \frac{1}{r^2}\;.
\end{align}
Let us compute the Yang-Mills equations in spherical coordinates. To
this end, note that the non-zero connection coefficients are 
\begin{align}
\Gamma^r_{ij} = -r\gamma_{ij}\;, \qquad \Gamma^i_{rj} = \frac{1}{r}\delta^i_j\;,\qquad  \Gamma^k_{ij} = \hat\gamma^{k}_{ij}\;,
\end{align}
where $\hat\gamma^{k}_{ij}$ are the connection coefficients for $\gamma_{ij}$. The Yang-Mills equations become
\begin{align}
 & D_r F_{ij} = D_i V_j - D_j V_i \cr
  & D_{[i}F_{jk]} = 0 \cr
  & D^iV_i = 0\cr
  & D_rV_i + \frac{2}{r}V_i + \frac{1}{r^2}D^jF_{ji} = 0\ ,
\end{align}
where $V_i = F_{ri}$ and $D^j$ is the gauge-covariant derivative on
$S^4$ (with indices raised by $\gamma^{ij}$).  

The simplest solution to these equations is to set $A_r=0$, $V_i=0$
and $\partial_rA_i=0$ so that $F_{ij}$ simply satisfies the Yang-Mills
equations on the four-sphere: $ D_{[i}F_{jk]}=0$ and $D^jF_{ji}=0$
with $F_{ij}$ independent of $r$. In fact this is exactly what we
want. Note that in Cartesian coordinates we have
\begin{align}
F_{ij} = \frac{\partial x^\mu}{\partial \theta^i}\frac{\partial x^\nu}{\partial \theta^j}F_{\mu\nu}\ .
\end{align}
Now the change of variables between $x^\mu $ and $r, \theta^i$ has the form:
\begin{align}
  x^0 &= r\cos\theta^1 \cr
  x^1 &= r\sin\theta^1\cos\theta^2 \cr
   &\vdots &\cr
x^{4}&= r\sin\theta^1\ldots\sin\theta^4  \;.
\end{align}
Thus
\begin{align}
F_{\mu\nu}\sim \frac{1}{r^2}\;,
\end{align}
as required.

Solutions to the above equations can be straightforwardly constructed
starting with the BPST instanton on $\mathbb R^4$ and
stereographically projecting to $S^4$; see \eg Appendix B of
\cite{Constable:2001ag}. Solutions for generic $\SU(N)$ gauge group
can then be obtained by replacing the Pauli matrices with $N \times N$
matrix representations of the $\mathfrak{su}(2)$ algebra, such that
$[T_i, T_j] = 2 i \epsilon_{ijk }T_k$. Let us point out two amusing
associated facts for the case of single instantons. First, the
solutions on $S^4$ satisfy
\begin{align}
  \label{eq:6}
  F \wedge F = \frac{8 \rho^4 ~\sum_{i = 1}^3 T_i^2}{\Big(1+\rho^2 + (1 -
  \rho^2)\cos{\theta^1}\Big)^4} 
  \sqrt \gamma ~ d^4 \theta\;.
\end{align}
Note that when $\rho = 1$, the coefficient collapses to $\frac{1}{2}$
and yields an $\SO(5)$-symmetric expression. Second, when the matrix
representation of $\SU(2)$ is irreducible
$\sum_{i = 1}^3T_i^2 = (N^2 -1) \one_{N \times N}$ and $F\wedge F$ is
gauge invariant. Upon using this one gets (for generic $\rho$)
\begin{align}
  \label{eq:7}
  I = \frac{1}{8 \pi^2} \Tr \int F \wedge F = \frac{N(N^2-1)}{6}\;,
\end{align}
where $\int d^4\theta \sqrt \gamma = 8 \pi^2/3$.  The above
ratio is always an integer and scales like $N^3$ for large $N$.

Yet another equivalent definition of instanton operators comes from
generalising the approach of \cite{Borokhov:2002ib}, that is by
requiring that $\cI_n(x)$ creates a charge-$n$ instanton-soliton in 5D
Yang-Mills theory. By definition this has $n$ units of instanton
charge. Due to the Bianchi identity for the gauge field, there is a
topological conserved current
\begin{align}
  \label{eq:1}
  J^\mu = \frac{1}{32\pi^2} \epsilon^{\mu\nu\kappa\lambda\rho}{\rm Tr}(F_{\nu\kappa}F_{\lambda\rho})\;.
\end{align}
The OPE of this current with $\mathcal I_n(0)$ is given by
\begin{equation}
  \label{eq:2}
  J^\mu(x) \cI_n(0) \sim \frac{3n}{8\pi^2} \frac{x^\mu}{|x|^5}\cI_n(0) + \cdots\;,
\end{equation}
with the ellipsis denoting less singular terms. The exact coefficient
can be deduced by requiring that the charge $I$ of the state obtained
by acting on the vacuum with ${\cal I}_n$ at $t=-\infty$ is $n$.

One could also introduce the notion of a `refined' instanton operator,
where in addition to specifying the topological instanton number on
the four-sphere one should also provide the moduli of the instanton on
$S^4$. Such operators would then not be Lorentz scalars as they will
not be rotationally invariant on the sphere.  However, we have no need
for these here and simply include an integration over all instanton
configurations at the insertion point.

\subsection{Supersymmetry and Supersymmetric States}

The supervariation of a fermion in the background of the Yang solution
is
\begin{align}\label{gaugino}
 \delta \psi =
\frac{1}{2}\Gamma^{\mu\nu}F_{\mu\nu}\Gamma_5 \varepsilon=
\frac{1}{2}\Gamma^{ij}F_{ij}\Gamma_5\varepsilon\;.  
\end{align}
Here we are using a convention where $\Gamma_5$ arises from the extra
dimension of the $(2,0)$ theory, which has been reduced on a
circle. For the maximally supersymmetric case we also need to impose
\begin{align}
\Gamma_{012345}\varepsilon=\varepsilon\ .\label{susycon}
\end{align}
We Wick rotate $x^0$, go to spherical coordinates \eqref{coords} and
introduce the frame
\begin{align}
e^{\underline r} = dr\qquad e^{\underline i} = r \tilde e^{\underline i}\;,
\end{align}
where $\tilde e^{\underline i}$ is a vielbein for $S^4$ with
unit radius.  The condition (\ref{susycon}) becomes
\begin{align}
\frac{1}{5!}\frac{1}{\det e} \epsilon^{m_1...m_5}\Gamma_{m_1}...\Gamma_{m_5}\Gamma_5\varepsilon =\frac{1}{4!}\frac{1}{r^4\det \tilde e} \epsilon^{m_1...m_4}\Gamma_{m_1}...\Gamma_{m_4}\Gamma_r\Gamma_5\varepsilon = i\varepsilon\ .
\end{align}
Note that we are still in flat Euclidean space.
Going to the vielbein frame we find
\begin{align}
 \Gamma_{\underline {1234}}\Gamma_{ r}\Gamma_5\varepsilon = i\varepsilon\ .
\end{align}
It is easy to see that upon imposing the above, along with the
selfduality condition obeyed by the background
\begin{align}
\frac{1}{2}\epsilon_{ijkl}F^{\underline {kl}} = \pm F_{\underline {ij}}\,
\end{align}
one can satisfy \eqref{gaugino} iff
\begin{align}
  \Gamma_{ r}\Gamma_5\varepsilon  = \mp i \varepsilon\;,
\end{align}
or equivalently
\begin{align}
  \label{eq:8}
\Big(  \frac{x^\mu}{|x|}\Gamma_\mu \Gamma_5 \pm i \Big) \varepsilon = 0\;,
\end{align}
where the signs are correlated.

Note that generically this would have to be true for all $x^\mu$,
which is impossible with a constant $\varepsilon$. To see this,
restrict $x^\mu$ to the $x^1$ axis. Then one concludes that
$\varepsilon$ has to be an eigenstate of
$\Gamma_1\Gamma_5$. Similarly, by restricting $x^\mu$ to the $x^2$
axis, one sees that $\varepsilon$ has to be an eigenstate of
$\Gamma_2\Gamma_5$. Given that $\Gamma_1\Gamma_5$ and
$\Gamma_2\Gamma_5$ do not commute, \eqref{eq:8} has no solution; in
other words, all supersymmetries are broken.
 
At first this may seem counter-intuitive: the theory on
$ {\mathbb R}^{1,4}$ contains solitonic BPS states which carry
instanton number and inserting an instanton operator at $t=-\infty$
should---by definition---create such a state out of the
vacuum. However, there is no direct contradiction. We remind the
reader that the 5D supersymmetry algebra is given by
\begin{align}
  \label{eq:3}
  \{Q_\alpha,Q_\beta\} =& P_\mu (\Gamma^\mu C^{-1})^-_{\alpha\beta} + Z_5 (\Gamma^5 C^{-1})^-_{\alpha\beta} +  Z_\mu^I (\Gamma^\mu\Gamma^I C^{-1})^-_{\alpha\beta} \cr& + Z_{5}^I (\Gamma^ 5\Gamma^I C^{-1})^-_{\alpha\beta}
+ Z_{\mu\nu\lambda}^{IJ}(\Gamma^{\mu\nu\lambda}\Gamma^{IJ}C^{-1})^-_{\alpha\beta}\;,
\end{align}
where we have taken the spinors to be those of eleven dimensions (\ie\
real with 32 components) with $C = \Gamma_0$ the charge conjugation
matrix defined by $\Gamma_M^T = -C\Gamma_M C^{-1}$,
$M=0,1,2,...,10$. We are using $x^5$ as the extra dimension associated
to M-theory; see \cite{Lambert:2010iw} for more details on
notation. The $Z_5$ central charge is in fact proportional to the
instanton number
\begin{align}
  \label{eq:4}
  Z_5 =- \frac{1}{2g^2_{YM}} {\rm Tr}  \int F\wedge F\;.
\end{align}
When the local instanton operator acts on the vacuum, it creates a
tower of states with different energies, all of which carry instanton
charge. The instanton-solitons with charge $n$ can be found at the
bottom of the tower by projecting out all other states
\begin{align}
  \label{eq:5}
  |n\rangle = \lim_{\tau\to\infty}e^{-(H -
    Z_5)\tau}\mathcal{I}_n(0) |0\rangle\;.
\end{align}
Once again, this is analogous to three dimensions, where monopole
operators and BPS vortices are annihilated by different
combinations of supercharges \cite{Intriligator:2013lca}.\footnote{See
  also \cite{Hayashi:2014hfa} for a similar discussion on operators and instanton-soliton states in the context of superconformal 5d $T_N$ theories.}

There exists an interesting exception to the above discussion: The
moduli space of instanton operators also includes configurations with
$\rho=0$. The corresponding classical solutions can be constructed
using singular (anti)instantons on the four-sphere. These have
$\delta$-function support on a single point of the $S^4$. In that case,
it is clear that \eqref{eq:8} need only be satisfied at that point,
setting half of the supersymmetry parameters to zero. As a result, the
$\rho=0$ instaton operators are $\frac{1}{2}$-BPS.\footnote{This argument
  did not appear in a previous version of this article. We would like
  to thank S.~Kim,  O.~Bergman and
  D.~Rodr\'iguez-G\'omez for   pointing it out to us.}

\subsection{Chern-Simons Terms} 
 
We next look at the effect of including Chern-Simons terms. Even
though these are excluded in parity-conserving theories, such as
maximally supersymmetric Yang-Mills, they can be important in other
contexts, such as $\cN = 1$ 5D gauge theories.

If the action also includes a  term
 \begin{align}\label{CSterm}
 S_{CS} = \frac{k}{24\pi^2}{\rm Tr}\int (F\wedge F\wedge A +\frac{i}{2}F\wedge A\wedge A\wedge A - \frac{1}{10}A\wedge A\wedge A\wedge A\wedge A)\;,
 \end{align}
which satisfies
\begin{align}
 \delta S_{CS} = \frac{k}{8\pi^2}{\rm Tr}\int F\wedge F\wedge \delta A  \;,
 \end{align}
then instanton operators are not gauge invariant. To see this, we
consider a gauge transformation $\delta A = D\omega$  where we assume
that $\omega=0$ at infinity. In this case, for  boundary conditions
corresponding to an instanton operator at position $x$, we
find\footnote{See \cite{Bagger:2012jb} for a similar discussion of
  monopole operators in 3D Chern-Simons theories.}
 \begin{align}
 \delta S_{CS} &= \frac{k}{8\pi^2}{\rm Tr}\int D\left(F\wedge F\wedge  \omega\right)  \nonumber\\
 & = \frac{k}{8\pi^2}{\rm Tr} \oint_{S_\infty^4} F\wedge F\omega  -\frac{k}{8\pi^2}{\rm Tr}\oint_{S_x^4}F\wedge F \omega  \nonumber\\
&= -\frac{k}{8\pi^2}{\rm Tr}\left[\omega(x)\oint_{S_x^4}F  \wedge
  F \right]\;.
\end{align}
 Assuming that the rest of the action and local operators are gauge invariant then
\begin{align}
 \delta \langle{\cal I}_n(x){\cal O}_{01}(x_1)\ldots& {\cal
   O}_{0k}(x_k)\rangle = \cr
&= -\int_{
\frac{1}{8\pi^2}{\rm Tr}\oint_{S^4_x} {F\wedge F}=n} [DXDAD\psi] \;{\cal O}_{01}(x_1)\ldots {\cal O}_{0k}(x_k) \delta S_{CS}e^{-S}\nonumber\\
 &=  \frac{k}{8 \pi^2}{\rm Tr}\left[\omega(x) \oint_{S_x^4}F\wedge F\right]\langle{\cal I}_n(x){\cal O}_{01}(x_1)\ldots {\cal O}_{0k}(x_k)\rangle \ .
 \end{align}
Thus to understand the gauge transformation properties of ${\cal I}_n$ requires knowing
\begin{equation}
Q_I = \frac{1}{8\pi^2} \oint_{S_x^4}F\wedge F 
\end{equation}
rather than just the instanton number $n = {\rm Tr}(Q_I)$. For the
single instanton irreducible case considered above in
\eqref{eq:6}, \eqref{eq:7} we have
\begin{align}
  \label{eq:9}
  Q_I = \frac{1}{6}\sum_{i = 1}^3 T_i^2\;,
\end{align}
which is independent of the moduli and leads to a gauge-invariant
instanton operator.

It would be interesting to examine
whether $Q_I$ plays a similar role to
\begin{equation}
Q_{M } = \frac{1}{2\pi} \oint_{S^2} F \ ,
\end{equation}
in the GNO analysis \cite{Goddard:1976qe}. This could lead to
instanton operators appearing in representations of the (dual) gauge
group.  To see some basic features of this quantity it is helpful to
introduce a basis $t_a$ of the full gauge group Lie algebra with
metric $\kappa_{ab} = {\rm Tr}(t_at_b) $ and symmetric tensor
\begin{equation}
d_{abc} =  {\rm Tr}(t_{(a}t_{b)}t_c)\ .
\end{equation}
In this case  
\begin{equation}
\delta {\cal I}_n = k d_{abc}  Q_{I}^{ab}  \omega^c {\cal I}_n\ .
\end{equation}
On the one hand, if the Lie algebra has an abelian ${\mathfrak u}(1)$ factor with
generator $t_0= \one_{N \times N}$, then under a gauge transformation of the form $\omega = \omega^0  t_0$ we see that
\begin{equation}
\delta {\cal I}_n = kn \omega^0 {\cal I}_n
\end{equation}
so that ${\cal I}_n$ carries $\U(1)$ charge $kn$.  On the other, if
the gauge group Lie algebra is ${\mathfrak su}(2)$, then $d_{abc}=0$
and hence ${\cal I}_n$ is gauge invariant, in accordance with what one
would expect from \eqref{eq:9} when $T_i = \sigma_i$.
 
We can also consider cases where there is a Chern-Simons term that
mixes the nonabelian gauge field with a background $\U(1)$ field $B$
(such as the one arising \eg in \cite{Kim:2012tr}):
\begin{align}
 S_{\U(1)\ CS} &=  \frac{k}{8\pi^2 }\int dB\wedge {\rm Tr} \left(F\wedge A+\frac{i}{3}A\wedge A\wedge A \right)\nonumber\\ 
 &= \frac{k}{8\pi^2 }\int B\wedge {\rm Tr} (F\wedge F )\ .
\end{align}
In this case, under a background gauge transformation $\delta B =
d\lambda$ and again with $\lambda=0$ at infinity,
\begin{align}
 \delta S_{\U(1)\ CS} = -\frac{k}{8\pi^2 }\oint_{S^4_x} \lambda(x) {\rm Tr} (F\wedge F ) = -kn\lambda(x)\;,
 \end{align}
where we have once again assumed the boundary conditions appropriate
for an instanton operator ${\cal I}_n(x)$. Therefore, as long as the
rest of the action is gauge invariant,
\begin{align}
  \delta \langle{\cal I}_n(x){\cal O}_{01}(x_1)\ldots& {\cal
                                                       O}_{0k}(x_k)\rangle
                                                       = \cr
  & = -\int_{
    \frac{1}{8\pi^2}{\rm Tr}\oint_{S^4_x} F\wedge F = n} [DXDAD\psi] \;{\cal O}_{01}(x_1)\ldots {\cal O}_{0k}(x_k) \delta S_{CS}e^{-S}\nonumber\\
                                                     & =  kn \lambda(x)\langle {\cal I}_n (x){\cal O}_{01}(x_1)\ldots {\cal O}_{0k}(x_k)\rangle
\end{align}
and an instanton operator ${\cal I}_n$ has background $\U(1)$ charge $
kn$.

If we think in terms of smooth soliton states on $\mathbb R^{1,4}$
that carry instanton number, then the effect of a Chern-Simons term
\eqref{CSterm} is to modify the equation of motion to
\begin{equation}
D_\mu F^{\mu\nu a} = -\frac{g^2 k}{32\pi^2}\kappa^{ad}d_{bcd}\varepsilon^{\nu\lambda\rho\sigma\tau}F_{\lambda\rho}^bF_{\sigma\tau}^c\ ,
\end{equation}
where $\kappa^{ab}$ is the inverse to the Lie-algebra metric $\kappa_{ab}$. This implies that
\begin{equation}
D_i F^{i 0 a} = \frac{g^2 k}{32\pi^2}\kappa^{ad}d_{bcd}\varepsilon^{ ijkl}F^b_{ij}F^c_{kl}\ ,
\end{equation}
so that the instanton acts as a source for the electric field-strength. 
 Thus the Chern-Simons term induces an electric charge as measured by the flux through the sphere at infinity
 \begin{align}
 \frac{1}{g^2}\int_{S^3_\infty} F^{i0}dS_i
 =  \frac{1}{g^2}\int D_iF^{i0} d^4x  = k d_{bcd}Q_{I}^{bc}\kappa^{ad}t_a\ ,
 \end{align}
where $Q_{I}$ is now evaluated as an integral over ${\mathbb R}^4$.
In particular, if the gauge group Lie algebra has a ${\mathfrak u}(1)$
generator $t^0 = \one_{N \times N}$, then 
 \begin{align}
 \frac{1}{g^2}{\rm Tr}\int_{S^4_\infty}   F^{i0}dS_i
 =    k n  \ .
 \end{align}
%

\section{An Extra Dimension and Enhanced Lorentz Symmetry}\label{applic}

As an SCFT with no lagrangian description, the six-dimensional $(2,0)$
theory can be captured completely by its spectrum and operator product
expansion coefficients. In relating this description to $\cN=2$ 5D
Yang-Mills, our first step is to explore what it means to compactify a
CFT on a circle and in turn what that implies for the correlations
functions.

Suppose that we have a CFT in six dimensions, consisting of a list of
local (gauge invariant) operators $\hat {\cal O }(x)$ as well as their
correlation functions
$\langle\hat {\cal O}_1(x_1)\ldots \hat {\cal O}_n(x_n)\rangle$. These
correlation functions are subject to the usual constraints of
conformal field theory; for example, for two operators of conformal
dimensions $\Delta_1$ and $\Delta_2$ we have
\begin{align}
\langle \hat {\cal O}_1(\hat x_1) \hat {\cal O}_2(\hat x_2)\rangle
= \frac{c_{12}}{|\hat x_1-\hat x_2|^{ {\Delta_1+\Delta_2} } }\ . 
\end{align}
Here and in what follows, we have used a hat to label all
uncompactified six-dimensional quantities.

We want to examine how these correlation functions behave once we
compactify one dimension. To this end, we let the six-dimensional
coordinates $\hat x$ be denoted by $(x,y)$ where $x$ is now a
five-vector and $y\in \mathbb R$. To compactify we view the circle as
an orbifold: $S^1 = {\mathbb R}/\Gamma$ where $\Gamma$ acts as
$(x,y)\to (x,y+2\pi R n)$, $n\in\mathbb Z$. Thus we could consider
operators of the form 
\begin{align} 
{\cal O}(x,y) := 
  \sum_{n\in
    \mathbb Z} \hat {\cal O}(x, y+2\pi R n) = \sum_{m\in\mathbb Z}
  e^{i my/R} {\cal O}_m(x)\;.
\end{align} 
We do not claim that all operators in the five- and six-dimensional
theories are related in this way. This will only apply to a special
class of operators such as BPS operators or ones which satisfy
linear equations of motion. For a study of such operators in the
setting of thermal CFT see \cite{Brigante:2005bq}. It is therefore
important to stress that we merely wish to consider correlation
functions involving operators of this form as an example.

In the above we introduced the Fourier modes
\begin{align}\label{fourmode}
{\cal O}_m(x) = \frac{1}{2\pi R
  }\int_0^{2\pi R}dy \; e^{-imy/R}\sum_{n\in \mathbb Z} \hat {\cal O}(x, y+2\pi R n) \ .
\end{align}
Clearly the ${\cal O}_m$ correspond to momentum eigenstates of operators around the $S^1$. From the five-dimensional perspective these are Kaluza-Klein modes.

Let us look at a generic two-point function on ${\mathbb R}^{5}\times S^1$:
\begin{align}
  \langle   {\cal O}_1(x_1,y_1)  {\cal O}_2(x_2,y_2)\rangle &=                                                               \sum_n\sum_m
                                                              \langle   \hat{\cal O}_1(x_1,y_1+2\pi Rn)  \hat{\cal O}_2(x_2,y_2+2\pi Rm)\rangle\cr
  &=
    \sum_n\sum_m
    \frac{c_{12}}{\Big(  x_{12}^2+( y_{12}+2\pi R(n-m))^2\Big)^{\frac{\Delta_1+\Delta_2}{2}}}
    \cr
  &=-\frac{1}{2}\sum_k
    \frac{c_{12}}{\Big( x_{12}^2+(y_{12}+ 2\pi  R k  )^2\Big)^{\frac{\Delta_1+\Delta_2}{2}}}\cr
  &=:\sum_{\ell } e^{i \ell (y_{12})/R} \Phi_{12\ell}(x_{12})\ ,
\end{align}
where $x_{12}$ is shorthand for $x_1 - x_2$ and in going from the
second to the third line we have used $\zeta$-function regularisation
when performing one of the sums, $\zeta(0) =
-\frac{1}{2}$.

We can also write
\begin{align}
\langle   {\cal O}_1(x_1,y_1)  {\cal O}_2(x_2,y_2)\rangle &= \sum_{n  }\sum_{m} e^{i (ny_1+my_2)/R}\langle {\cal O}_n(x_1){\cal O}_m(x_2)\rangle \cr
&= \sum_{n  }  e^{i n y_{12}/R}\langle {\cal O}_n(x_1){\cal O}_{-n}(x_2)\rangle\ ,
\end{align}
where in the last line we have used translational invariance, which implies
\begin{align}
\langle {\cal O}_n(x_1){\cal O}_{m}(x_2)\rangle =0\quad{\rm if}\quad n\ne -m\ .
\end{align}
Matching these two expressions gives
\begin{align}
  \Phi_{12n}(x_{12}) &= \langle {\cal O}_n(x_1){\cal
    O}_{-n}(x_2)\rangle \cr \nonumber &= -\frac{1}{4\pi R }\int_0^{2\pi
    R} dy_{12} e^{- in y_{12}/R} \sum_k \frac{c_{12}}{\Big(x_{12}^2+(y_{12}+ 2\pi Rk)^2\Big)^{\frac{\Delta_1+\Delta_2}{2}}}\;.
 \end{align}
To proceed we can evaluate the sum using
\begin{align}
 \frac{1}{\Big(x_{12}^2+(y_{12}+ 2\pi Rk)^2\Big)^{s}}
 = \frac{ \pi^{s}} {\Gamma(s)} \int_0^\infty \frac{dt}{t^{1+s}}  e^{-\frac{\pi}{t}(x_{12}^2+(y_{12}+ 2\pi Rk)^2) }\ ,
\end{align}
as well as Poisson resummation:
\begin{align}
\sum_m e^{-\pi A(m+a)^2 } = \sum_m A^{-1/2} e^{-\pi A^{-1}m^2 - 2\pi im a}
\ .
\end{align}
This gives
\begin{align}
\langle {\cal O}_n(x_1){\cal O}_{-n}(x_2)\rangle &=-\frac{c_{12}
  \pi^{\frac{\Delta_1+\Delta_2}{2}}}{8\pi^2 R^2
  \Gamma(\frac{\Delta_1+\Delta_2}{2})}\sum_m \int_0^{2\pi R}
\int_0^\infty   \frac{dy_{12}dt}{t^{1
    +\frac{\Delta_1+\Delta_2-1}{2}}}e^{- in y_{12}/R}\cr &
\times e^{-\frac{\pi}{t}x_{12}^2-   \frac{t} {4\pi R^2} m^2-
  iy_{12}m/R }\cr
 &= -\frac{c_{12} \pi^{\frac{\Delta_1+\Delta_2}{2}}}{4\pi  R
   \Gamma(\frac{\Delta_1+\Delta_2}{2})}  \int_0^\infty   \frac{
   dt}{t^{1 +\frac{\Delta_1+\Delta_2-1}{2}}} e^{-\frac{\pi}{t} x_{12}^2-   \frac{t} {4\pi R^2} n^2 }\ .
\end{align}
For $n=0$ we simply find
\begin{align}
\langle {\cal O}_0(x_1){\cal O}_{0}(x_2)\rangle = -\frac{c_{12}\sqrt{\pi}}{4\pi R}\frac{\Gamma(\frac{\Delta_1+\Delta_2-1}{2} )}{\Gamma( \frac{\Delta_1+\Delta_2}{2})}\frac{1}{|x_{12}|^{ {\Delta_1+\Delta_2 -1} }}\ .
\end{align}
However, for the Kaluza-Klein modes we obtain
\begin{align}
\langle {\cal O}_n(x_1){\cal O}_{-n}(x_2)\rangle = -\frac{c_{12} \pi^{\frac{\Delta_1+\Delta_2}{2}}}{2 \pi  R  \Gamma(\frac{\Delta_1+\Delta_2}{2})}\left( \frac{|n|}{2\pi R|x_{12}|}\right)^{\frac{\Delta_1+\Delta_2-1}{2}}K_{\frac{\Delta_1+\Delta_2-1}{2}}\left(\frac{|n||x_{12}|}{ R}\right)\ .
\end{align}
Here we have used the integral expression for a Bessel function:
\begin{align}
\int_0^\infty \frac{d t}{t^{1+s}} e^{-at -b/t}=2\left|\frac{a}{b}\right|^{\frac{s}{2}} K_{s}(2\sqrt{ab})\;.
\end{align}

Let us now make the further identification, valid for the case of an M5-brane wrapped on a circle, that $R  = g^2/4\pi^2$, where $g^2$ is the five-dimensional Yang-Mills coupling constant. We see that
\begin{align}
\langle {\cal O}_0(x_1){\cal O}_{0}(x_2)\rangle = -\frac{ c_{12}\pi^\frac{3}{2}}{g^2}\frac{\Gamma(\frac{\Delta_1+\Delta_2-1}{2} )}{\Gamma( \frac{\Delta_1+\Delta_2}{2})}\frac{1}{|x_{12}|^{ {\Delta_1+\Delta_2 -1} }}\ 
\end{align}
and
\begin{align}
  \langle {\cal O}_n(x_1){\cal O}_{-n}(x_2)\rangle &=  - \frac{2\pi c_{12} \pi^{\frac{\Delta_1+\Delta_2}{2}}}{ g^2 \Gamma(\frac{\Delta_1+\Delta_2}{2})}\left( \frac{2\pi |n|}{g^2|x_{12}|}\right)^{\frac{\Delta_1+\Delta_2-1}{2}}K_{\frac{\Delta_1+\Delta_2-1}{2}}\left(\frac{4\pi^2}{g^2}|n||x_{12}| \right)\\
                                                   &=-\frac{c_{12}\pi^{\frac{\Delta_1+\Delta_2}{2}}}{ 2|n| \Gamma(\frac{\Delta_1+\Delta_2}{2})} \left(\frac{2\pi|n| }{g^2|x_{12}|}\right)^{\frac{\Delta_1+\Delta_2}{2}} e^{-\frac{4\pi^2}{g^2} |n||x_{12}| }\left( 1+{\cal O}\Big(\frac{g^2}{|n||x_{12}|}\Big) \right)\nonumber\ , \end{align}
where we have expanded out the Bessel function for small $g^2$  using
\begin{align}
K_s(z) = \sqrt\frac{\pi}{2z}e^{-z}(1+\ldots )\ .
\end{align}
Thus we see that $\langle {\cal O}_0(x_1){\cal O}_{0}(x_2)\rangle$ has
a purely perturbative interpretation in the five-dimensional gauge
theory but $\langle {\cal O}_n(x_1){\cal O}_{-n}(x_2)\rangle$ is
non-perturbative. In particular, it carries the distinctive
exponential dependence  $e^{-S_n}$ on the coupling $g$, where
\begin{align}\label{Sclassical}
S_{n} = \frac{4\pi^2}{g^2} |n||x_1-x_2|\ .
\end{align}

\subsection{Matching Kaluza-Klein Modes to Instanton Operators}

To capture these correlators from a five-dimensional viewpoint let us define, for any
zero-mode operator ${\cal O}_0(x)$ constructed out of local
five-dimensional fields (not necessarily gauge invariant),
\begin{align}\label{IOdef}
{\cal O}_n(x) := {\cal I}_n(x){\cal O}_0(x)\;.
\end{align}
First, note that this is consistent with the fact that instanton
operators are not supersymmetric in 5D. Suppose we start with a BPS
operator $\hat {\cal O}$ in six dimensions. This implies that there is
a supercharge $ Q$ such that $[ Q, \hat {\cal O}]=0$. Let us 
introduce a superspace with Grassmaniann coordinate $\theta$ in such a
way that $Q$ is realised as
\begin{align}
 Q = \frac{\partial}{\partial \theta} + i\bar \theta\Gamma^m \partial_m\ . 
\end{align}
Then using \eqref{fourmode} one obtains
\begin{align}
[ Q,{\cal O}_n] = \frac{n}{2\pi R^2}\bar \theta\Gamma^y {\cal O}_n \ne 0\ .
\end{align}
Therefore, while the Kaluza-Klein zero-modes for BPS operators are
still BPS, their associated higher Fourier modes are not.\footnote{In
  the identification of 5D super-Yang-Mills with the compactification
  of the (2,0) theory on a circle, the supercharges for the two
  theories remain the same.}  This seems intuitively clear since
taking a BPS state at rest and adding momentum (but not boosting it)
will violate the BPS saturation condition.\footnote{However, a way
  around this argument that could be relevant to the zero-sized
  instanton case discussed above, is to modify the definition of the
  Fourier modes by changing the exponent from $e^{-imy/R}$ to
  $e^{-im(y+i\bar\theta \gamma^y\theta)/R}$ in \eqref{fourmode}.}

Next, we show that the definition of ${\cal O}_n(x)$ \eqref{IOdef} leads to
momentum conservation along the  $S^1$. To do this we note that a single-sourced Yang field configuration has
\begin{align}
S &= \frac{1}{4g^2} {\rm Tr}\int d^5xF_{\mu\nu}F^{\mu\nu}\nonumber\\
 &= \frac{1}{4g^2} {\rm Tr}\int_0^\infty dr~ r^4 \oint_{S^4}d\Omega_4 \frac{F_{ij}F^{ij}}{r^4}\nonumber\\
 &= \frac{4\pi^2|n|}{g^2}\int_0^\infty   dr\ ,
\end{align}
which is finite near $r=0$ but diverges as $r=R\to\infty$ like
\begin{align}
S = \frac{4\pi^2|n|}{g^2}R\ .\label{InstAct}
\end{align}
Thus $S\to\infty$ and the path integral vanishes. However, if we have
a correlation function where two or more instanton operators are
inserted, we can then obtain a finite action if the total instanton
number is zero; \ie if $ {\rm Tr}\oint_{S^4_\infty} {F\wedge F}=0$ where $S^4_\infty$ is the four-sphere at infinity. Therefore
\begin{align}
\langle{\cal O}_{n_11}(x_1){\cal O}_{n_22}(x_2)...{\cal O}_{n_kk}(x_2)\rangle =0\ ,
\end{align}
unless 
\begin{align}
\sum_{i=1}^kn_i=0\ .
\end{align}

This is consistent with momentum conservation along the $S^1$ and
crucially different from monopole operators in the M2-brane
interpretation. The latter are used to create eleven-dimensional
momentum, where the momentum is off the M2-brane and hence not
conserved. Here we wish to construct momentum states along the
M5-brane and hence require conservation of momentum.

Finally, let us show that the correlation function computed in 5D
Yang-Mills reproduces the $e^{-S_{n}}$ dependence that we saw in
Eq.~\eqref{Sclassical}.  The evaluation of a correlation function
involving insertions of two or more instanton operators is dominated
by the action of a classical solution that satisfies the boundary
conditions
\begin{align}
\frac{1}{8\pi^2} {\rm Tr}\oint_{S^4_i} F\wedge F = n_i 
\end{align}
at each of the insertion points $x_i$. However, to construct such a solution seems very difficult. 

Instead, consider the case of two instanton operator insertions located at $x_1$ and $x_2$ in five-dimensional Euclidean space, with instanton charges $n$ and $-n$ respectively. Based on dimensional analysis,  the minimum action for a field configuration is 
\begin{align}
S_{min} = \frac{1}{4g^2}\int d^5x  F_{\mu\nu}F^{\mu\nu}= \frac{K}{g^2}|x_1-x_2|\ ,
\end{align}
where $K$ is a dimensionless constant. To determine $K$ we can consider the case where one instanton operator is at $ x_1 = 0$ and the other is sent to infinity: $|x_2|=R\to\infty$. The diverging action will then be dominated by a single instanton operator of charge $n$ located at $ x_1= 0$. Comparing with (\ref{InstAct}) we see that $K = 4\pi^2 |n|$ and hence  
\begin{align}
S_{min} = \frac{4\pi^2|n|}{ g^2}|x_1-x_2| = S_{n}\ .
\end{align}
This reproduces the correct dependence of instanton-operator
contributions to the path integral, as one would expect for the
compactification of six-dimensional correlators on $S^1$ from \eqref{Sclassical}.

\section{Summary}\label{summary} 

In this note we have discussed a particular class of disorder
operators in five-dimensional gauge theories, dubbed instanton
operators. These are defined through a modification of the boundary
conditions for the gauge field in the path integral, which imposes a
non-vanishing second Chern class on any four-sphere that surrounds the
insertion point (but no insertion point of other instanton operators
in the same correlation function) in Euclidean space. We have examined
various properties of these operators---such as the fact that they are generically
not BPS---and in particular we argued that they can be identified as
inserting discrete units of six-dimensional momentum into maximally
supersymmetric five-dimensional Yang-Mills. Therefore, they play an
important role in enhancing the Lorentz symmetry to $\SO(1,5)$,
relating the theory to the six-dimensional $(2,0)$ SCFT and hence in
providing a UV completion.

It would be very interesting to see if these or similar operators also
have a role to play in minimally supersymmetric five-dimensional
Yang-Mills theories, perhaps leading instead to enhanced global
symmetries \cite{Seiberg:1996bd, Morrison:1996xf, Seiberg:1997ax,
  Intriligator:1997pq}; for related recent work see
\cite{Bergman:2012kr, Kim:2012gu, Jafferis:2012iv, Bashkirov:2012re,Bergman:2013koa,
  Bergman:2013ala, Bergman:2013aca, Bergman:2014kza, Mitev:2014jza}.

\section*{Acknowledgements} 

We would like to thank S.~Cremonesi, A.~Kapustin, K.~Papadodimas,
G. Papadopoulos, A.~Royston and S.~Ramgoolam for helpful discussions
and comments. N.L. is supported in part by STFC grant
ST/J002798/1. C.P. is a Royal Society Research Fellow. We would also
like to acknowledge CERN where some of this work was carried out.

\bibliographystyle{utphys}
\bibliography{Yang}

\end{document}